\documentclass{aa}  

\usepackage{graphicx}
\usepackage{txfonts}
\usepackage{hyperref}
\usepackage{xcolor}

\def\Msun{\mbox{${\rm M}_{\odot}$}}

\def\Rsun{\mbox{${\rm R}_{\odot}$}}

\def\kms{\mbox{km\,s$^{-1}$}}

\newcommand{\spec}[3]{\mbox{#1\,{\sc #2}\,$\lambda$#3}\xspace}

\begin{document}

   \title{VLTI/GRAVITY enables the determination of the first dynamical masses of a classical Be + stripped and bloated pre-subdwarf binary}

   \author{R.\ Klement
          \inst{1}
          \and
          Th.\ Rivinius
          \inst{1}
          \and
          D.\ Baade
          \inst{2}
          \and
          A.\ Mérand
          \inst{2}
          \and
          J.\ Bodensteiner
          \inst{2}
          \and
          A.\ J.\ Frost
          \inst{1}
          \and
          H.\ Sana 
          \inst{3,4}
          \and
          T.\ Shenar 
          \inst{5}
          \and
          D.\ R.\ Gies 
          \inst{6}
          \and
          P.~Hadrava 
          \inst{7}
          }

   \institute{European Organisation for Astronomical Research in the Southern Hemisphere (ESO), Casilla 19001, Santiago 19, Chile\\
              \email{robertklement@gmail.com}
        \and
             European Organisation for Astronomical Research in the Southern Hemisphere (ESO), Karl-Schwarzschild-Str.\ 2,\\ 85748 Garching bei M\"unchen, Germany
        \and Institute of Astronomy, KU Leuven, Celestijnlaan 200D, 3001 Leuven, Belgium \and
Leuven Gravity Institute, KU Leuven, Celestijnenlaan 200D, box 2415, 3001 Leuven, Belgium
        \and  The School of Physics and Astronomy, Tel Aviv University, Tel Aviv 6997801, Israel
        \and Center for High Angular Resolution Astronomy, Department of Physics and Astronomy,\\ Georgia State University, P.O. Box 5060, Atlanta, GA 30302-5060, USA
        \and Astronomical Institute, Academy of Sciences of the Czech Republic, Bo\v{c}n\'{\i} II 1401, CZ-14100 Prague, Czech Republic
             }



  \abstract
   {HR~6819 is the first post-mass transfer binary system composed of a classical Be star and a bloated pre-subdwarf stripped star directly confirmed by interferometry. While the Be star is already spun up to near-critical rotation and possesses a self-ejected viscous Keplerian disk, the stripped star is found in a short-lived evolutionary stage, in which it retains the spectral appearance of a B-type main-sequence star while contracting into a faint subdwarf OB-type star.}
   {In order to understand the evolution of intermediate-mass interacting binaries, the fundamental parameters of cornerstone objects such as HR~6819 need to be known. We aim to obtain orbital parameters and model-independent dynamical masses of this binary system to quantitatively characterize this rarely observed evolutionary stage.}
   {We analyzed a time series of 12 interferometric near-IR $K$-band observations from VLTI/GRAVITY with the help of the geometrical model-fitting tool PMOIRED. We included recently published radial velocities based on FEROS high-resolution spectroscopy for the binary orbital solution.
   }
   {With the GRAVITY data, we obtained the astrometric orbit, relative fluxes of the components, and parameters of the circumstellar disk of the Be star; we also detected helium line signatures from the stripped star. Using the published radial velocities enabled us to obtain the dynamical masses of the components as well as the dynamical parallax. The Be star is the slightly brighter component in the $K$ band and is almost 16 times as massive as the bloated stripped star, with the individual dynamical masses being $4.24\pm0.31$\,{\Msun} for the Be star and $0.270\pm0.056$\,{\Msun} for the stripped star. The orbit is slightly eccentric, with $e=0.0289\pm0.0058$, and the semimajor axis of the orbit is $0.3800\pm0.0093$\,AU. The distance derived from the orbital solution is $296.0\pm8.0$\,pc, significantly lower than the distance from \textit{Gaia} DR3, which is overestimated by $\sim$24\% due to the orbital motion.}
   {The newly obtained fundamental parameters provide an important anchor for evolutionary models of interacting binaries and for the physics of mass transfer. The low mass of the bloated star means that it may become completely undetectable once it settles into a faint subdwarf, which implies that many more Be stars may have low-mass companions despite appearing single.}

               
\titlerunning{Dynamical masses of classical Be + bloated stripped star binary}
   \maketitle
%

\section{Introduction} \label{sec:introduction}

HR~6819 (HD~167128, QV~Tel; $G=5.3$) is among the first discovered binary systems composed of a classical Be star and a bloated pre-subdwarf (pre-sd) stripped star \citep{2020A&A...641A..43B, 2021MNRAS.502.3436E}, and the first whose nature was unambiguously confirmed by optical interferometry \citep{2022A&A...659L...3F}. This binary system is a product of recent mass transfer between two B-type components, which originally orbited each other on a close orbit before the more massive component evolved to fill its Roche lobe. Most of the mass of this component, currently observed as the pre-sd star, was lost to its companion, which was rejuvenated and spun up to near-critical rotation in the process. After the mass transfer ceased, the spun-up component formed a viscous Keplerian disk, thus becoming a classical Be star \citep{2013A&ARv..21...69R}. The pre-sd star is currently found in a short-lived evolutionary stage that lasts only a few hundred thousand years \citep{2021MNRAS.502.3436E}, which precedes its settling into a much fainter helium-burning subdwarf OB-type (sdOB) star and the spectrum becoming dominated by the Be star. HR~6819 is of great importance to stellar astrophysics in general and to the evolution of binary stars in particular as it represents a newly identified, brief snapshot in the evolution of interacting binaries \citep{2024ARA&A..62...21M}. The immediate post-mass-transfer stage of HR~6819 also offers a window into the physics of the mass-transfer process, the surface chemical composition of the newly formed Be star, and the surface chemical composition of the exposed core of the former mass donor. 

HR~6819 belongs among the so-called black hole (BH) impostors, which are multiple stellar systems that were initially erroneously claimed to contain dormant (non-accreting) stellar mass BHs \citep{2022Msngr.186....3B}. HR~6819 specifically was reported to be a hierarchical triple system \citep{2020A&A...637L...3R}, only to be shown later to consist solely of a classical Be star and a bloated pre-sd star of similar brightness, orbiting each other on a $\sim$40-day orbit \citep{2020A&A...641A..43B, 2020ApJ...898L..44G, 2021MNRAS.502.3436E}. Several other systems similar to HR~6819 have been interferometrically confirmed more recently \citep{2024arXiv241209720R}, including another BH impostor and the first suggested Be + pre-sd binary, LB-1 \citep{2019Natur.575..618L,2020Natur.580E..11A,2020A&A...639L...6S}.

The key to the unambiguous confirmation of the nature of HR~6819 and similar systems is the employment of high-angular-resolution observations. They include the single-telescope techniques of speckle interferometry and adaptive-optics-assisted imaging and, more importantly, multiple-telescope optical/near-IR interferometry \citep{2023ARA&A..61..237E}. In the triple-star BH scenario, two widely separated, similarly bright components would be resolved, while in the binary pre-sd scenario, no such outer component would be seen. Despite the tentative report of a speckle companion with unconstrained brightness \citep{2021ATel14340....1K}, no bright wide companion was detected with follow-up observations by VLT/MUSE, thus confirming the Be + pre-sd scenario \citep{2022A&A...659L...3F}. Furthermore, given the proximity and brightness of HR~6819, optical interferometry obtained with the Very Large Telescope Interferometer (VLTI) was able to resolve the two binary components themselves, enabling the detection of the orbital motion on the basis of two observations separated by $\sim$13 days \citep{2022A&A...659L...3F}.  

Binary systems such as HR~6819 are cornerstone objects for our understanding of the formation of classical Be stars \citep{2014ApJ...796...37S,2021ApJ...908...67S,2021A&A...653A.144H} and rapid rotators in general \citep{2013ApJ...764..166D}. Be stars represent up to 40\% of B-type stars depending on the metallicity of the environment, and several observational clues indicate that they may be dominated by descendants of HR~6819-like systems. Most apparent is the lack of close, main-sequence companions to (early-type) Be stars \citep{2020A&A...641A..42B}, and the majority of Be stars thus appear to be single. When companions are found, they are either evolved products of stripping in mass-transferring binaries, such as neutron stars in Be X-ray binaries \citep{2011Ap&SS.332....1R}, or their nature remains uncertain \citep[e.g.,][]{2000ASPC..214..668G}. The hardest to detect and characterize are stripped sdOB and (pre-)white dwarf (WD) companions, although they are expected to be particularly common if most Be stars originate from mass transfer in binaries \citep{1991A&A...241..419P,2014ApJ...796...37S}. Currently, there are only $\sim$25 confirmed Be + sdOB binaries \citep{2018ApJ...853..156W,2021AJ....161..248W,2022ApJ...940...86K}, and there are only candidate Be + WD systems, among them the class of peculiar X-ray sources with the prototype $\gamma$~Cas \citep{2016AdSpR..58..782S,2022MNRAS.510.2286N,2023ApJ...942L...6G}.

Other observational clues suggest that many Be stars may possess low-mass companions at small orbital separations. Among these, the prevalence of spectral energy distribution (SED) turndown (i.e., the steepening of the spectral slope toward radio wavelengths observed in all Be stars with sufficient radio data) is indicative of disk truncation possibly caused by low-mass companions \citep{1991A&A...244..120W, 2019ApJ...885..147K}. Furthermore, the structure of the radio SED for the several Be stars with radio SED measurements at multiple wavelengths, including the known binary $\gamma$~Cas, suggests the presence of a circumbinary extension of the Be disk \citep{2017A&A...601A..74K}. In several Be stars, signs of circum-companion gas --- sometimes in the form of a disk --- were observed as well \citep[e.g.,][de Amorim et al., in prep.]{2002ApJ...573..812B,2018ApJ...865...76C,2024ApJ...962...70K}, including in HR~6819 itself \citep{2020A&A...641A..43B} as well as in LB-1 \citep{2020A&A...639L...6S,2022A&A...660A..17H}. Recent smoothed particle hydrodynamic simulations of Be disks in binary systems strongly support the notion laid out above, that the Be disks are not truncated by the binary orbit but instead form complex structures, including a bridge region connecting the inner Be disk to the companion, circum-companion structures that may take the form of an accretion disk, and a large-scale circumbinary spiral arm that may contribute to the radio fluxes \citep{rubio_et_al}.

The aim of this work is to advance our knowledge of HR~6819 and its place among classical Be stars by obtaining model-independent dynamical masses of the binary system. They were derived with the help of a time series of high-angular-resolution and high-spectral-resolution interferometric observations from VLTI/GRAVITY, which are described in Sect.~\ref{sec:observations}. The fitting of the relative astrometry of both components, as well as of the Be star and its disk parameters, is described in Sect.~\ref{sec:fitting}, while the combined three-dimensional orbital solution using previously published radial velocity (RV) measurements from high-resolution FEROS spectroscopy is outlined in Sect.~\ref{sec:orbital_solution}. The results are discussed in Sect.~\ref{sec:discussion} in the context of our current understanding of the origin and the general properties of classical Be stars.


\section{VLTI/GRAVITY observations}\label{sec:observations}

A total of 12 snapshots (1 hr observing time including calibrator data and overheads) of HR~6819 were collected between 2021 September and 2023 September by the VLTI \citep{2023ARA&A..61..237E} four-telescope beam combiner GRAVITY \citep{2017A&A...602A..94G} operating in the near-IR $K$ band ($1.98$ to $2.40$\,$\mu$m). The 1.8 m Auxiliary Telescope (AT) Array, with each AT using the dedicated adaptive optics system NAOMI \citep{2019A&A...629A..41W}, was used in the ``Large'' configuration (maximum baseline of $\sim$130\,m) in order to achieve the highest angular resolution that was possible at VLTI at the time ($\lambda / 2B_{\rm max} \sim 1.75$\,mas in the near-infrared $K$ band).

The GRAVITY instrument was used in the single-field on-axis mode, which splits the light equally between the fringe tracker \citep{2019A&A...624A..99L} and the science beam combiner. The spectral resolution of the science spectrograph was set to high ($\lambda / \Delta \lambda \sim 4000$), enabling a detailed analysis of spectro-interferometric features. Owing to its brightness, HR~6819 was observed in relaxed atmospheric conditions (seeing $<1.4''$ and coherence time in the V band $\tau_0 > 1.4$\,ms). The data were reduced and calibrated using the official European Southern Observatory pipeline version 1.6.6 in the EsoReflex environment \citep{2013A&A...559A..96F}. This resulted in the following observables covering the entire $K$ band for each snapshot: spectrum, four closure phases (T3PHI, one per telescope triangle), six absolute visibilities (|V|) and six differential phases (DPHI, one of each per telescope pair). 

Each observation of HR~6819 consisted of an object--sky--object sequence, resulting in two sets of data points, which may differ from each other due to the changes in baseline projection caused by the Earth rotation (particularly in |V|; see Fig.~\ref{fig:bestfit}). Every observation of HR~6819 was followed by an observation of the calibrator HD~161420 ($K = 5.4$, uniform disk diameter $=0.321\pm0.008$\,mas, spectral type F0\,IV) in order to enable the absolute calibration of |V| and T3PHI. The calibrator diameter is small considering the angular resolution of the data and the calibrator data thus represent the instrumental response of a point-like source, which is ideal for calibration purposes. The first two calibrated snapshot observations from 2021 were taken in the split polarimetric mode with a Wollaston prism in the beam path, while the rest of the data were obtained in the combined mode. The GRAVITY data (Table~\ref{tab:astrometry}) are of very good quality, although for the snapshot taken on 2023~May~01, we had to discard the |V| part of the dataset due to a miscalibration issue (calibrated $\mathrm{|V|} >1$), probably caused by rapidly changing atmospheric conditions. The full set of the reduced and calibrated data will be made available in the Optical Interferometry Database\footnote{\url{http://oidb.jmmc.fr/index.html}} \citep{2014SPIE.9146E..0OH}.

To visualize and model the reduced GRAVITY data, we used the Python code Parametric Modeling of Optical InteRferomEtric Data\footnote{\url{https://github.com/amerand/PMOIRED}} \citep[PMOIRED;][]{2022SPIE12183E..1NM}. PMOIRED enables the fitting of multicomponent geometrical models composed of, for example, uniform disks (UDs) and Gaussians, to interferometric data, including the possibility to fit normalized line profiles and associated spectro-interferometric signatures when the spectral resolution is sufficient. PMOIRED uses least-squares minimization to obtain best-fit parameters, and contains features such as telluric correction of the $K$-band spectrum for GRAVITY data, grid search for the detection of binary companions, and a bootstrapping resampling method to obtain realistic uncertainties of the fitted parameters. The grid search method is the same as implemented in the CANDID\footnote{\url{https://github.com/amerand/CANDID}} code dedicated specifically to interferometric companion detection \citep{2015A&A...579A..68G}.

The fitting of spectro-interferometric features enables the derivation of RVs, whose accuracy depends on the wavelength calibration of the data. The accuracy of GRAVITY in the high spectral resolution mode is known to be at least $\sim0.02$\% \citep{2023A&A...672A.119G}, with the pipeline delivering the data at vacuum wavelengths. To check the wavelength calibration precision for each snapshot, we used PMOIRED to fit and subtract telluric features from the $K$-band spectrum. This was done by interpolating a one-dimensional grid of Molecfit \citep{2015A&A...576A..77S} models parameterized by the precipitable water vapor while correcting for the shape of the continuum. This procedure resulted in a telluric-corrected $K$-band spectrum (tellurics do not affect the interferometric observables), the fitted precipitable water vapor, and finally the calibrated spectral dispersion model, which was then applied to the data to facilitate the highest possible accuracy of the fitted RVs. The accuracy of the wavelength calibration resulting from the spectral dispersion model is plotted in Fig.~\ref{fig:wavcal} for each snapshot, revealing that it is better than expected and no worse than $0.004$\% ($\sim$12\,\kms) for all epochs except for 2021~September~05, where it is $0.007$\% ($\sim$21\,\kms). To obtain realistic uncertainties of the fitted RVs from GRAVITY data (Sect.~\ref{sec:GRAVITY_RVs}), we thus conservatively chose to quadratically add the uncertainty of $12$\,{\kms} to all epochs except for 2021~September~05, for which we added the uncertainty of $21$\,{\kms}.

%

\section{Geometrical model fitting}\label{sec:fitting}

\subsection{Relative astrometry and flux ratios of the two binary components}\label{sec:astrometry}

The relative astrometry and flux ratios of the two components were derived using geometrical model fitting with the PMOIRED code to the spectral region around the hydrogen Br$\gamma$ line, which is the most prominent spectral line in the $K$ band. We selected a 16\,nm-wide spectral region centered at the Br$\gamma$ vacuum wavelength of $\lambda_{\mathrm{Br}\gamma\mathrm{, vac}} = 2.1661178$\,$\mu$m, and normalized the line profile and DPHI using the region edges. Br$\gamma$ shows a double-peaked emission line profile compatible with what is expected from a Keplerian disk surrounding the Be star and seen at intermediate inclination (see Fig.~\ref{fig:bestfit}). The spectro-interferometric features in Br$\gamma$ (T3PHI, DPHI, and |V|) contain a combination of the signal from the Keplerian Be disk, and a much more pronounced signal from the photocenter displacement caused by the bright companion. The companion itself does not appear to show any features in the Br$\gamma$ region. We fitted the normalized Br$\gamma$ line profile, and the interferometric observables T3PHI, DPHI, and |V|, simultaneously.

The first two GRAVITY snapshots taken in 2021 were already analyzed by \citet{2022A&A...659L...3F}, first using a simple model approximating the Br$\gamma$ line profile with a Gaussian, and subsequently with a geometrical model mimicking a more realistic contribution from the Be star disk with receding and approaching spatially offset components, which result in a double-peaked line profile similar to the observed one. Our results for the relative astrometry for these two epochs, using a more complex geometrical model (see below), are consistent with these preliminary results.

To obtain a realistic representation of the Be star disk in our geometrical model, we used the prescription for a Keplerian disk that was recently implemented in PMOIRED \citep{2011A&A...529A..87D, 2012A&A...538A.110M,2024ApJ...962...70K}. The full model used to obtain the final fits of the Br$\gamma$ spectral region consists of the following components: UDs for the two stellar components (with diameters UD$_{\rm Be}$ and UD$_{\rm pre-sd}$) with a relative astrometric offset ($\Delta {\rm RA}$, $\Delta {\rm Dec}$ of the pre-sd star relative to the Be star), and with a relative flux ratio (parameterized as the contribution of the pre-sd star to the total continuum flux $f_{\rm pre-sd}$), and a Keplerian disk whose position coincides with that of the Be star. The parameters of the Keplerian disk, contributing solely in the spectral line, are the following: the angular full width at half maximum of the Br$\gamma$ line emission region (FWHM$_{\mathrm{disk, Br}_\gamma}$), disk inclination ($i_{\rm disk}$, assuming a geometrically thin disk so that the flattening ratio is $r=\cos{i_{\rm disk}}$), disk position angle (PA$_{\rm disk}$, measured from north to east), the inner diameter of the disk, the Keplerian rotational velocity $v_{\rm orb}$ at the inner disk diameter (re-parameterized through $v_{\rm orb} \sin{i_{\rm disk}}$), the power law exponent of the radial rotation law ($\beta$), the central wavelength of the disk Br$\gamma$ signature (re-parameterized as RV of the Be star RV$_{\rm Be}$), and the equivalent width of the Br$\gamma$ line profile (EW$_{\mathrm{Br}\gamma}$). All parameters were kept free in the fitting process with the following exceptions: both UD$_{\rm Be}$ and UD$_{\rm pre-sd}$ were fixed at $0.15$\,mas (see below), the inner diameter of the Keplerian disk was kept equal to UD$_{\rm Be}$, and $\beta$ was fixed at $0.5$ to represent Keplerian rotation. The remaining free parameters were therefore $v_{\rm orb} \sin{i_{\rm disk}}$, RV$_{\rm Be}$, $\Delta {\rm RA}$, $\Delta {\rm Dec}$, $f_{\rm pre-sd}$, FWHM$_{\mathrm{disk, Br}_\gamma}$, $i_{\rm disk}$, PA$_{\rm disk}$, and EW$_{\mathrm{Br}\gamma}$.

As for the choice of the UD values of both components, we ran an initial fit with the value of $0.12$\,mas, corresponding to the absolute radii given by \citet{2020A&A...641A..43B} and \citet{2021MNRAS.502.3436E} scaled to their adopted distances. As the distance resulting from the dynamical parallax of our orbital fits (Sect.~\ref{sec:combined_orbital_solution}) is significantly lower than the adopted distances in the previous works, we repeated the fitting process with the value of $0.15$\,mas for the UDs, which resulted from rescaling the radii to the smaller distance. We note that the exact values of the UDs do not influence the astrometric results, as in the considered range they represent angularly unresolved components at the angular resolution of the data. However, the value of UD$_{\rm Be}$ does influence the results for the Keplerian disk, as it corresponds to the disk's inner diameter. Namely, the higher value of UD$_{\rm Be}$ results in slightly lower fitted values of the Keplerian velocity at the base of the disk, which is in our model given by $v_{\rm orb} \sin{i_{\rm disk}}$. As we could not fit the value of UD$_{\rm Be}$, the exact choice of the best-estimated value of $0.15$\,mas thus introduces a possible systematic bias into the fitted parameters of the Keplerian disk.

The model including the Keplerian disk described above successfully converged for all epochs with $\chi^2_{\rm red.}$ ranging from $1.8$ to $5.3$, and with the resulting astrometric positions showing a clear orbital motion according to the expectations. A representative example of one of the datasets and the quality of our model fit is shown in Fig.~\ref{fig:bestfit}, along with synthetic images and component fluxes for the resulting geometrical model shown in Fig.~\ref{fig:image}. The final values of the fitted parameters and their uncertainties were determined with a bootstrapping algorithm implemented in PMOIRED, the results of which are shown in Fig.~\ref{fig:bootstrap} for the same representative epoch. The relative positions and flux ratios for each epoch are listed in Table~\ref{tab:astrometry}. We note that the errors do not include contributions from possible systematics such as small-scale miscalibration of the data due to rapidly changing atmospheric conditions during observations.

The weighted average of the flux ratio is $f_{\rm pre-sd} = 0.439\pm0.013$, that is to say, the pre-sd star is the slightly fainter component, contributing $\sim$44\% of the total $K$ band continuum flux on average (the total flux includes a possible small contribution from the Be star disk). Some variability in the flux ratio is apparent with the minimum and maximum values from individual snapshots being $0.376\pm0.018$ and $0.540\pm0.028$, respectively, although most measurements cluster around the average value. The resulting parameters for the Be star and its disk are given in Tables~\ref{tab:other_params_average} and \ref{tab:other_params} and are discussed in Sect.~\ref{sec:disk}.

\subsection{The Be star and its Keplerian disk}\label{sec:disk}

\begin{table}[]
\caption[xx]{\label{tab:other_params_average} Average disk parameters from the fit to GRAVITY data.}
\begin{center}
\begin{tabular}{lcc}
\hline\hline
Parameter  & AVG$_{\rm weighted}$ (1$\sigma$)\tablefootmark{a} & Full range\tablefootmark{b} \\
\hline
$v_{\rm orb} \sin{i_{\rm disk}}$ [\kms]           &  $242\pm12$ & [225, 303]\\
$i_{\rm disk}$ [$^\circ$]     & $28.9\pm5.7$ & [19.6, 40.9]\\
PA$_{\rm disk}$ [$^\circ$]    & $27.9\pm5.5$ & [3.4, 66.7] \\
FWHM$_{\mathrm{Br}\gamma\mathrm{, disk}}$ [mas] & $0.481\pm0.046$ & [0.375, 0.621]    \\
EW$_{\mathrm{Br}\gamma\mathrm{, disk}}$ [nm] & $0.702\pm0.052$ & [0.556, 0.999]   \\
\hline
\end{tabular}
\tablefoot{
\tablefoottext{a}{Weighted averages across the individual epochs with 1$\sigma$ uncertainties.}
\tablefoottext{b}{Taken from the probability distributions, an example of which is plotted in Fig.~\ref{fig:bootstrap}}
}
\end{center}
\end{table}

\begin{figure*}[]
\begin{center}
\includegraphics[width=17cm]{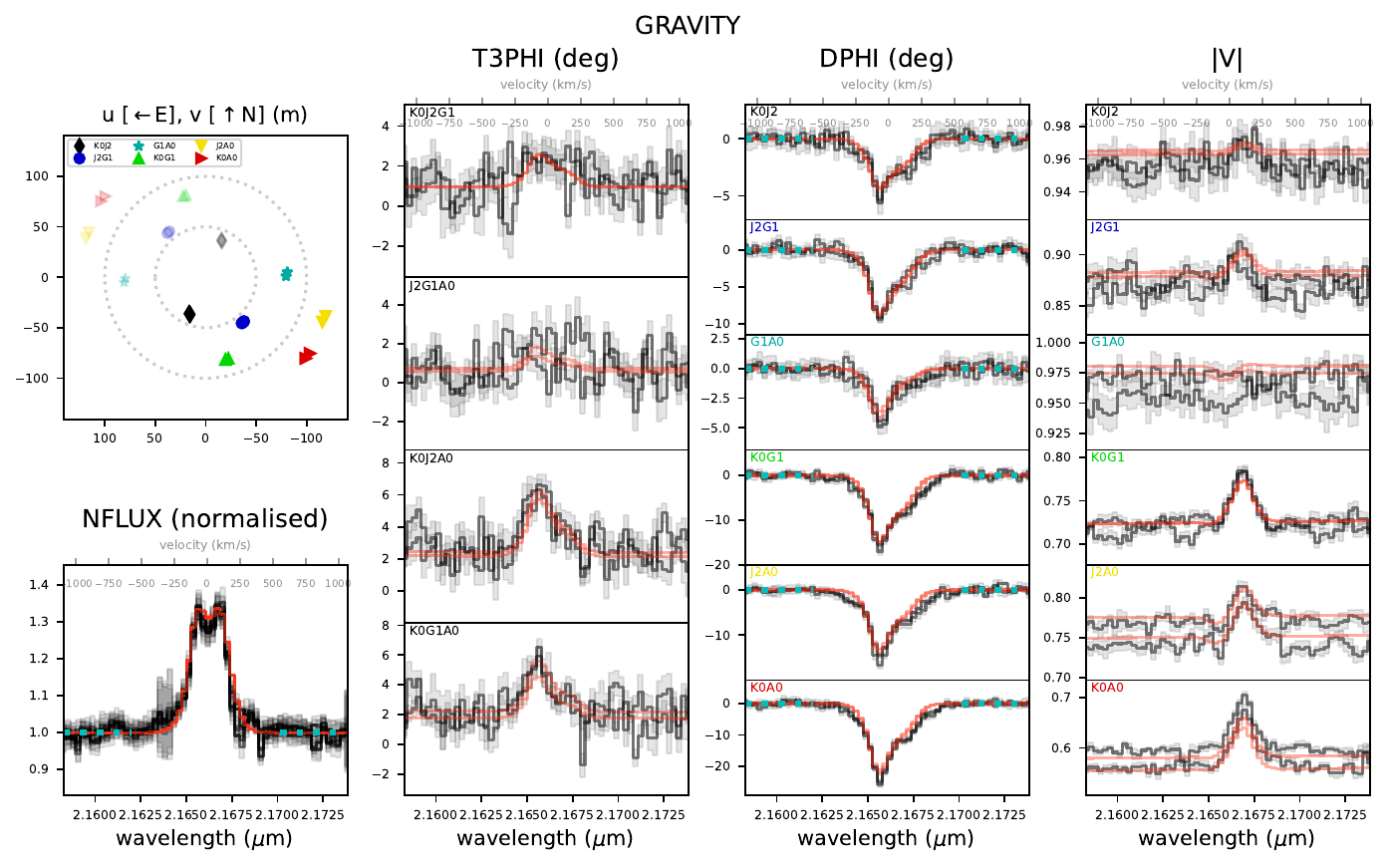}%
\end{center}
\caption[xx]{\label{fig:bestfit} PMOIRED Keplerian geometrical model fit to the calibrated GRAVITY snapshot taken on 2023~July~17, representing the average quality of the model fit ($\chi^2=3.1$) as well as the typical quality of the data. The upper-left panel shows the $(u,v)$ coverage, the lower-left panel the normalized Br$\gamma$ line profile (NFLUX), and the remaining panels the interferometric observables closure phase (T3PHI), differential phase (DPHI), and the absolute visibility (|V|). The data (black with gray error bars) are overplotted with the model fit (red). The blue dots in the NFLUX and DPHI panels represent the continuum region used for normalization. The baselines and telescope triangles are given in the corner of each panel, and their colors correspond to the $(u,v)$ coverage panel. For the layout of the VLTI telescope locations, see Fig.~10 in the VLTI manual (\url{https://www.eso.org/sci/facilities/paranal/telescopes/vlti/documents/VLT-MAN-ESO-15000-4552_v115.pdf}).
}
\end{figure*}

\begin{figure*}[]
\begin{center}
\includegraphics[width=17cm]{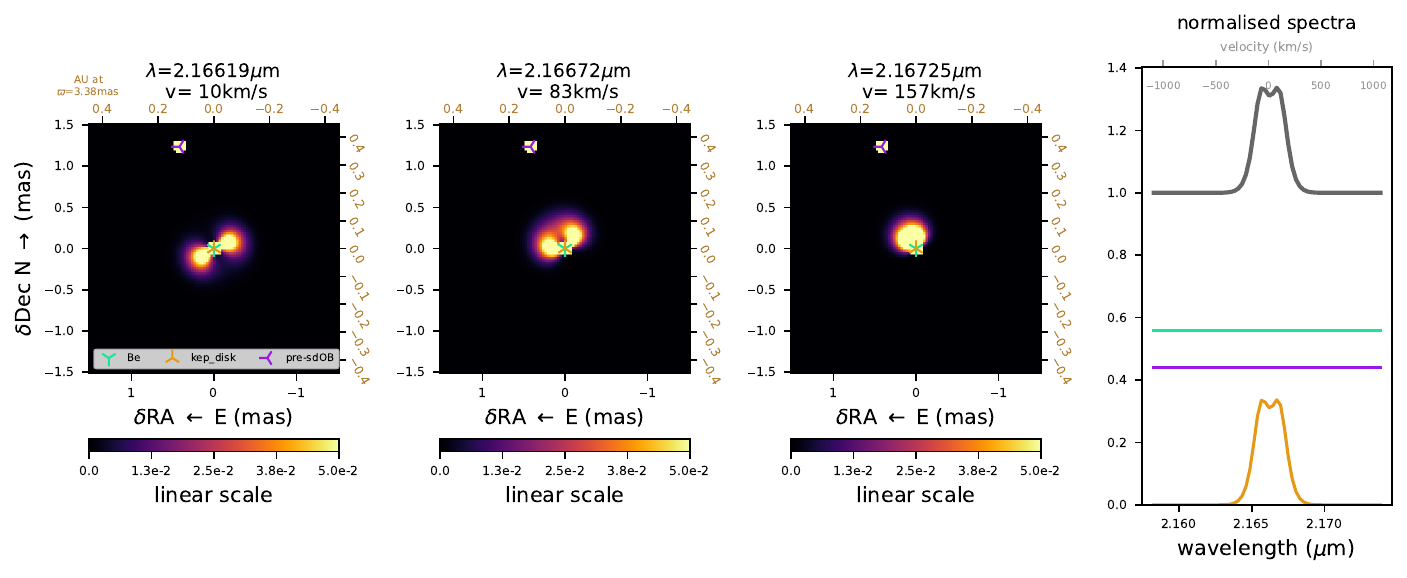}%
\end{center}
\caption[xx]{\label{fig:image} PMOIRED synthetic image resulting from the Keplerian model fit to the 2023~July~17 snapshot (left) and the fluxes of the model components (right). The model components are the Be star primary (green) and Keplerian disk (orange) in the image center, and the pre-sd secondary (purple) offset in the image by $\sim1.3$\,mas toward the northeast, with the sum of their fluxes normalized to $1.0$ in the continuum (black).
}
\end{figure*}

The measurement of the Keplerian disk parameters is complicated by the fact that the spectro-interferometric signals in Br$\gamma$ (T3PHI, DPHI, and |V|) are dominated by the photocenter displacement caused by the bright companion rather than by the typical signature of a resolved Keplerian disk (see Fig.~\ref{fig:kepDisk_only}). It also needs to be considered that the changing $(u,v)$ coverage of the data across the epochs means that they do not constrain the parameters in exactly the same way. Nevertheless, the overall consistency of the results across the epochs (listed in Table~\ref{tab:other_params}) as well as the generally very good quality of the fit to the observed data (Fig.~\ref{fig:bestfit}) shows that we are partly resolving the Br$\gamma$ line-emitting region of the Keplerian disk, even though its size is clearly close to the limits of the angular resolution of the data. The resulting Keplerian disk parameters are given in Table~\ref{tab:other_params_average}, which lists the weighted average of the values across all epochs as well as their full range, including errors.

The resulting value of $v_{\rm orb} \sin{i_{\rm disk}}$ is dependent on the fixed value of UD$_{\rm Be}=0.15$\,mas \citep[corresponding to Be star radius $R_{\rm Be} \sim 4.7$\,{\Rsun} at the distance of $296$\,pc; see Sect.~\ref{sec:combined_orbital_solution} and][]{2021MNRAS.502.3436E}. Relying on this value of UD$_{\rm Be}$, we can estimate the critical rotation fraction of the Be star from comparison with a previously derived $v_{\rm rot} \sin{i}$, where $v_{\rm rot}$ is the surface rotational velocity of the Be star, while assuming that $i=i_{\rm disk}$ (i.e., that the rotational planes of the Be star and its disk are aligned). The spectroscopic analyses of both \citet{2020A&A...641A..43B} and \citet{2021MNRAS.502.3436E} arrived at a consistent value of $v_{\rm rot} \sin{i} \sim 180$\,\kms. Adopting the larger uncertainty of $\pm20$\,{\kms} from \citet{2021MNRAS.502.3436E}, the resulting rotational parameter $W=v_{\rm rot}/v_{\rm orb}$ is $0.744\pm0.091$ (corresponding to $v_{\rm rot}/v_{\rm crit}=0.81\pm0.08$ where $v_{\rm crit}$ is the critical rotation velocity). This value is typical for classical Be stars, which generally rotate close to but below their critical velocity \citep[][]{2013A&ARv..21...69R}.

The remaining parameters of the Keplerian disk are realistic, albeit not particularly precise, given their spread across the individual epochs. Still, they provide a consistency check against the orbital solution, as for a disk in a post mass-transfer binary, $i_{\rm disk}$ should be equal to the orbital inclination $i$, while PA$_{\rm disk}$ should be parallel to the line of nodes given by $\Omega$. While the average PA$_{\rm disk}$ is indeed equal to $\Omega$, the average $i_{\rm disk}$ is $\sim2\sigma$ below $i$ (Sect.~\ref{sec:orbital_solution}, Table~\ref{tab:other_params_average}). Inspection of the individual epochs in which the resulting $i_{\rm disk}$ is significantly below $i$ (these include the 2023~July~17 epoch shown in Fig.~\ref{fig:bestfit}) reveals that the corresponding best-fit models appear to underestimate the central reversal of the Br$\gamma$ line profile in each case. On the other hand, for the epochs when $i_{\rm disk}$ is consistent with $i$, the Br$\gamma$ central reversal is deeper and more consistent with the observations. Thus, we conclude that the disagreement between the average $i_{\rm disk}$ and the orbital $i$ is not significant and the overall results are consistent with the Be disk being aligned with the orbit. We note that fixing the values of $i_{\rm disk}$ at different values in the full range (Table~\ref{tab:other_params_average}) during the fitting process has negligible influence on the resulting astrometric positions and flux ratios.

%
\section{Orbital solution for HR~6819}\label{sec:orbital_solution}

\subsection{Astrometric orbital solution}

\begin{table}[]
\caption[xx]{\label{tab:astrometric_solution} Purely astrometric orbital solutions.}
\begin{center}
\begin{tabular}{lcc}
\hline\hline
Parameter  & Circular             & Eccentric \\
\hline
$P$ [d]     & $40.305\pm0.012$   & $40.329\pm0.011$ \\
$T_0$ [RJD]     & $60073.36\pm0.33$     & $60076.41\pm0.97$ \\
$e$        & $0.0$                   & $0.0306\pm0.0071$   \\
$a''$ [mas]      & $1.290\pm0.019$ & $1.288\pm0.015$\\
$i$ [$^\circ$]        & $38.8\pm1.4$ & $39.5\pm1.0$\\
$\Omega$ [$^\circ$]   & $23.6\pm3.1$ & $23.1\pm2.1$ \\ 
$\omega_{\rm pre-sd}$ [$^\circ$]   & $90$ & $117.6\pm8.9$\\
\hline
rms residual [$\mu$as] & $25$ & $19$\\
\hline
\end{tabular}
\end{center}
\tablefoot{$P$ is the orbital period, $T_0$ is the epoch of periastron or the epoch of the superior conjunction, $e$ is the eccentricity, $a''$ is the angular semimajor axis of the orbit, $i$ is the orbital inclination, $\Omega$ is the longitude of the ascending node, $\omega$ is the longitude of the periastron (fixed at $90^{\circ}$ for circular orbits).}
\end{table}

\begin{figure}[t]
\begin{center}
\includegraphics[width=\hsize]{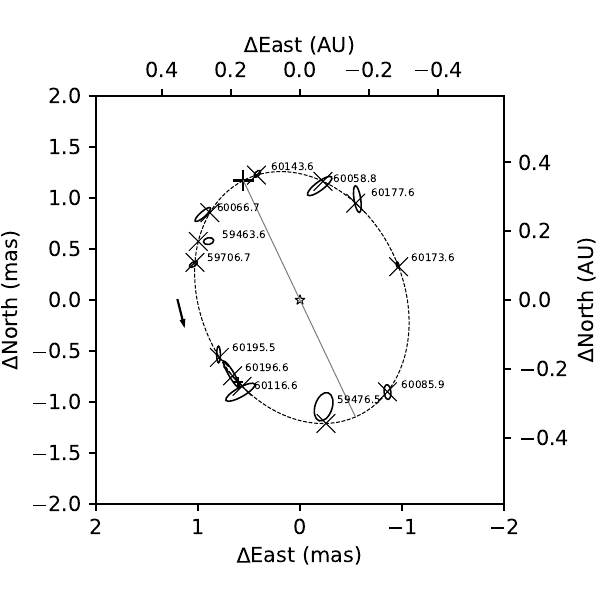}%
\end{center}
\caption[xx]{\label{fig:astrometric_orbit} Astrometric orbit for the eccentric combined solution using Be star RVs (dashed line) with the Be star at the center (star symbol), the line of nodes (gray line), the ascending node (large plus sign), and the periastron position (small plus sign coinciding with the astrometric position on HJD=60116.6). The relative astrometric positions of the pre-sd star (error ellipses corresponding to 3$\sigma$ uncertainties) coinciding with the calculated positions on the orbit (x symbols) are shown alongside the average HJD of each observation. For the secondary axes in units of AU, a distance of $296$\,pc is assumed.
}
\end{figure}

\begin{figure}[t]
\begin{center}
\includegraphics[width=\hsize]{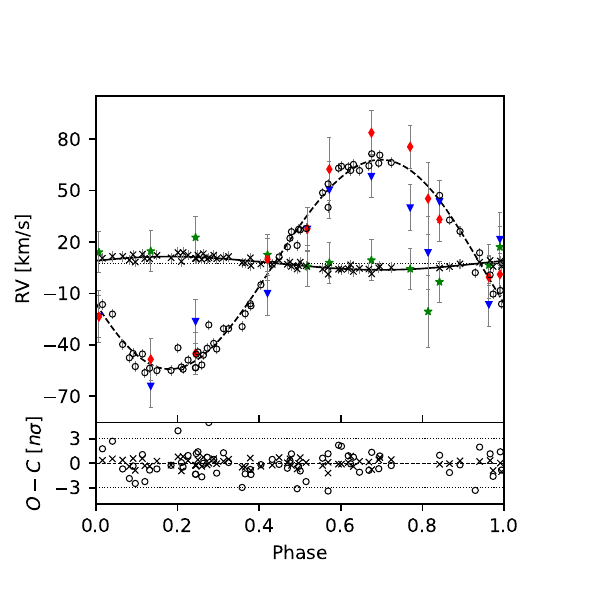}%
\end{center}
\caption[xx]{\label{fig:RV_curve} RV curves for the combined eccentric solution using Be star RVs. The Be star and pre-sd star RV curves (solid and dashed curves, respectively) are overplotted with the RVs measured by \citet{2020ApJ...898L..44G} for the Be star and the pre-sd star (x symbols and empty circles, respectively), RVs measured from the Keplerian disk signature in Br$\gamma$ in GRAVITY data (green stars), and RVs from \spec{He}{i}{20587} and \spec{He}{i}{21126} DPHI signatures in GRAVITY data (blue triangles and red diamonds, respectively). The errors of the RVs derived from GRAVITY data are dominated by the wavelength calibration uncertainty (Sect.~\ref{sec:observations}). The lower panel shows the residuals in units of $\sigma$ for the main RV dataset of \citet{2020ApJ...898L..44G}.
}
\end{figure}

To fit the astrometric positions in order to obtain the purely astrometric as well as the combined orbital solutions (Sect.~\ref{sec:combined_orbital_solution}), we used the IDL code orbfit-lib\footnote{\url{https://www.chara.gsu.edu/analysis-software/orbfit-lib}} and the implemented Newton-Raphson method, which performs a formal orbital fit by calculating a first-order Taylor expansion for the equations of orbital motion to minimize the residuals \citep{2016AJ....152..213S}. Unlike recent studies concerning HR~6819 \citep{2020A&A...641A..43B, 2020ApJ...898L..44G, 2021MNRAS.502.3436E}, we considered the Be star as the primary component, as it is both (slightly) brighter in the $K$ band on average (Sect.~\ref{sec:astrometry}) and currently much more massive than the pre-sd star.

The 12 sets of relative astrometry determined above were used to derive astrometric orbital solutions, assuming both circular and eccentric orbits. In both solutions (as well as in the combined solution derived in Sect.~\ref{sec:combined_orbital_solution}; see Fig.~\ref{fig:astrometric_orbit}), there are three astrometric points that do not fit the orbit within $3\sigma$. One of these is the lower quality dataset from 2023~May~01, which may be affected by a miscalibration issue that had forced us to completely discard the |V| part of that dataset (Sect.~\ref{sec:observations}). The other two are the datasets obtained in 2021 in split polarimetric mode, but it is unclear if any possible bias introduced by the polarimetric mode could be responsible for the lower quality of the orbital fit. To test how these three astrometric points influence the orbital solutions, we tried excluding them from the orbital fits, but the resulting parameters remained well within their respective $1\sigma$ errors in each individual solution.

The eccentric orbital solution results in considerably smaller residuals, with $\chi^2_{\rm red.}$ decreasing from $10.6$ to $5.5$. Although the obtained eccentricity is small at $e = 0.0306\pm0.0071$, it appears to be significant relative to its uncertainty at a $3\sigma$ level, as well as according to the statistical F-test comparing the circular and eccentric solutions. Nevertheless, with the exception of $e$ and the associated epoch and longitude of the inferior conjunction or periastron, respectively, for the circular and eccentric solutions ($T_0$ and $\omega_{\rm pre-sd}$), the resulting orbital parameters are compatible within 1$\sigma$ between the two solutions. The results are also all consistent with the previous works that analyzed FEROS spectra \citep{2020A&A...637L...3R, 2020A&A...641A..43B, 2021MNRAS.502.3436E}, while the angular semimajor orbital axis ($a''$), and the position angle of the line of nodes ($\Omega$) are both in a rough agreement with the preliminary values determined by \citet{2022A&A...659L...3F}. The orbital inclination $i\sim39^{\circ}$ resulting from our solution lies in between the values estimated indirectly by \citet{2020A&A...641A..43B} and \citet{ 2021MNRAS.502.3436E} and that of the preliminary astrometric fit of \citet{2022A&A...659L...3F}. The resulting orbital elements for the purely astrometric solution are listed in Table~\ref{tab:astrometric_solution}.

\subsection{Combined orbital solution}\label{sec:combined_orbital_solution}

The secondary pre-sd star presents a multitude of narrow absorption lines in its optical spectrum, which enables accurate and precise measurements of its RV \citep[e.g.,][their Fig.~7]{2020A&A...641A..43B}. The pre-sd RVs were recently measured from archival FEROS spectra with different procedures on four separate occasions, leading to consistent results with the velocity semi-amplitude $K_{\rm pre-sd} \sim 61$\,\kms \citep{2020A&A...637L...3R, 2020A&A...641A..43B, 2020ApJ...898L..44G, 2021MNRAS.502.3436E}. For our combined orbital solution, we specifically utilized the RVs from \citet{2020A&A...641A..43B}, obtained by fitting several lines simultaneously with Gaussian profiles, and RVs from \citet{2020ApJ...898L..44G}, obtained from cross-correlation with model spectra. The resulting orbital elements ended up being consistent within the respective 1$\sigma$ errors when using either of these RV sets, with the only exception being the systemic velocity, which has no bearing on the physical parameters. The final combined solutions presented in Table~\ref{tab:combined_solution} were obtained using the pre-sd RVs from \citet{2020ApJ...898L..44G} since they contain one additional measurement from an archival IUE spectrum, which helped increase the time baseline. 

As for RV measurements of the Be star, the situation is complicated by the rotationally broadened and disk-emission-contaminated spectral lines, as well as the extreme mass ratio resulting in a very small semi-amplitude $K_{\rm Be}$. From the recent works, only \citet{2020ApJ...898L..44G} published measurements for the individual FEROS epochs, obtained from fitting the broad wings of the H$\alpha$ emission line, which originate from the very inner parts of the Be star disk, and therefore trace the radial motion of the Be star itself. Using the RV curve of the pre-sd star, \citet{2020ApJ...898L..44G} further refined the Be star RV star measurements by first subtracting the estimated photospheric profile of the pre-sd star from the composite H$\alpha$ profiles at each epoch, resulting in the RV semi-amplitude $K_{\rm Be} = 3.9\pm0.7$. This value is consistent within $1\sigma$ with those obtained by \citet[][$K_{\rm Be} = 4.0\pm0.8$]{2020A&A...641A..43B} and \citet[][$K_{\rm Be} = 4.5\pm1.0$]{2021MNRAS.502.3436E} using different methods. We used both $K_{\rm Be}$ and the individual Be star RVs from \citet{2020ApJ...898L..44G} in subsequent combined orbital solutions (see below). 

For the combined orbital solution, the uncertainties of the astrometric and RV datasets were scaled in order to balance their contribution to the overall residuals. We considered both circular and eccentric orbits, similar to the purely astrometric solution described above. For the first set of combined solutions, we fitted the astrometry and pre-sd RVs, and used $K_{\rm Be}$ to calculate the masses and the distance (first and third columns in Table~\ref{tab:combined_solution}). In a slightly different approach, we also fitted the astrometry along with the RVs for both components (second and fourth columns in Table~\ref{tab:combined_solution}). These two approaches returned consistent results, as expected, with the only difference being the distribution of the resulting parameter uncertainties. Specifically, when fitting RVs of both components, the resulting $K_{\rm pre-sd}$ has a significantly smaller uncertainty than given by \citet{2020ApJ...898L..44G}, while for most of the other parameters, the uncertainties are slightly higher. This then results in a better-constrained $M_{\rm pre-sd}$, and slightly higher uncertainties for the other computed parameters. As for the purely astrometric solution, allowing for nonzero eccentricity results in overall lower residuals, with the eccentricity remaining small but significant at $e\sim0.029$. Both the masses and the distance are slightly higher in the eccentric solution than in the circular solution, but remain consistent within their respective errors. The astrometric orbit and RV curve corresponding to the eccentric solution using Be star RVs (fourth column in Table~\ref{tab:combined_solution}) are plotted in Figs.~\ref{fig:astrometric_orbit} and \ref{fig:RV_curve}, respectively.

The final results are listed in Table~\ref{tab:combined_solution}. Given the significance of the small but nonzero eccentricity, we took only the eccentric solutions into account for our final values and obtain $d=296.0\pm8.0$\,pc, $M_{\rm Be}=4.24\pm0.31$\,{\Msun}, $M_{\rm pre-sd}=0.270\pm0.056$\,\Msun, and $q=M_{\rm pre-sd}/M_{\rm Be} = 0.064\pm0.012$ ($M_{\rm Be}/M_{\rm pre-sd} = 15.6\pm3.0$). If we considered more conservatively the full error ranges from both the circular and eccentric solutions, the resulting parameters would be $d=291\pm13$\,pc, $M_{\rm Be}=4.12\pm0.43$\,{\Msun}, $M_{\rm pre-sd}=0.265\pm0.061$\,\Msun, and the same $q$ as given above for the eccentric solutions only. 

The distance resulting from the dynamical parallax of $296.0\pm8.0$\,pc is significantly different from the \textit{Gaia} DR3 distance of $368\pm16$\,pc \citep[calculated from the measured parallax while using photogeometric priors;][]{2021AJ....161..147B}. Without the knowledge of the Be star RVs, using the \textit{Gaia} distance to compute $K_{\rm Be}$ would result in $18$\,{\kms} or more for the four orbital solutions in Table~\ref{tab:combined_solution}, a value that is clearly overestimated by more than a factor of 4. This confirms that \textit{Gaia} DR3 parallaxes can be strongly biased by binary motions such as those of HR~6819. HR~6819 was not included among non-single stars in the recent release of the \textit{Gaia} DR3 catalog \citep{2023A&A...674A...1G,2023A&A...674A..34G}. However, this is not surprising given the stringent acceptance criteria for astrometric binary star processing in the DR3 release \citep{2023A&A...674A...9H}, and both the relative error of the measured parallax and the orbital period of HR~6819 being rather low.

\subsection{Radial velocities from GRAVITY data}\label{sec:GRAVITY_RVs}

Radial velocity measurements for the Be star disk, which should trace the Be star itself, are a direct by-product of the fitting of the Keplerian disk signature to the Br$\gamma$ region described in Sect.~\ref{sec:fitting}. Unfortunately, the weakness of the Be disk signal (see Sect.~\ref{sec:disk}) limits the precision of these RVs, despite the high wavelength calibration accuracy of the data. Nevertheless, the Br$\gamma$ RVs do follow the Be star orbit determined from the combined solution, but with an rms error of $\sim$9\,{\kms}, which is higher than the actual $K_{\rm Be}$ of $\sim$4\,\kms.

We searched the entire $K$-band spectral range of the GRAVITY data for any additional spectral features that would enable us to obtain further information about the binary system. We found weak and narrow DPHI signatures in two helium lines \spec{He}{i}{20587} (Fig.~\ref{fig:HeI20587}) and \spec{He}{i}{21126}. While the \spec{He}{i}{20587} line profiles show a hint of possible emission at $<$5\% of the continuum level, those of \spec{He}{i}{21126} appear to show an absorption profile at a similar level. The DPHI signature is clearer for \spec{He}{i}{21126}, while at certain epochs it is barely discernible in \spec{He}{i}{20587}. Fitting of Gaussian profiles to the two lines shows that they originate from (the vicinity of) the pre-sd star rather than from the disk of the Be star. The weakness of the DPHI signal limits the accuracy and precision of the extracted RVs, but they still clearly trace the RV curve of the pre-sd star (Fig.~\ref{fig:RV_curve}). The rms residual is $\sim$19\,{\kms} for the weaker \spec{He}{i}{20587}, and $\sim$10\,{\kms} for \spec{He}{i}{21126}.

If the \spec{He}{i}{20587} is indeed in emission (Fig.~\ref{fig:HeI20587}), we might be detecting small amounts of circumstellar gas surrounding the pre-sd star \citep[cf.][]{1996ApJS..107..281H}, while the \spec{He}{i}{21126} line probably originates in the photosphere of the pre-sd star. Given that the Be star is already surrounded by a well-developed self-ejected disk, parts of which might be reaching all the way to the secondary pre-sd star \citep{2020ApJ...898L..44G}, any circumstellar material around the secondary probably originates from accreted material from the Be disk. A possible presence of circum-companion gas was already suggested by \citet{2020A&A...641A..43B} based on a small emission peak in H$\alpha$ apparently following the orbit of the stripped star. The presence of an extended ``bridge'' region connecting the Be disk to the companion, as well as the possible presence of circum-companion structure formed by accreted material, is in line with recent predictions based on state-of-the-art hydrodynamical modeling \citep{rubio_et_al}.

\begin{table*}[]
\caption[xx]{\label{tab:combined_solution} Combined orbital solutions and dynamical masses.
}

\begin{center}
\begin{tabular}{lcccc}
\hline\hline
 &  \multicolumn{2}{c}{circular solutions} &  \multicolumn{2}{c}{eccentric solutions} \\
  & astro + RV${_{\rm pre-sd}}$ + $K_{\rm Be}$ & astro + RVs & astro + RV${_{\rm pre-sd}}$ + $K_{\rm Be}$ & astro + RVs \\
\hline
$P$ [d]                           & $40.3270\pm0.0012$  & $40.3266\pm0.0016$   & $40.32640\pm0.00094$ & $40.3261\pm0.0013$ \\
$T_0$ [RJD]                       & $53117.276\pm0.073$ & $53117.295\pm0.094$  & $53220.49\pm0.74$    & $53220.6\pm1.0$    \\
$e$                               & $0.0$               & $0.0$                & $0.0288\pm0.0043$    & $0.0289\pm0.0058$    \\
$a''$ [mas]                       & $1.298\pm0.013$     & $1.299\pm0.017$      & $1.283\pm0.011$      & $1.284\pm0.015$    \\
$i$ [$^\circ$]                    & $39.86\pm0.84$      & $39.9\pm1.2$         & $39.38\pm0.69$       & $39.37\pm0.95$       \\
$\Omega$ [$^\circ$]               & $26.0\pm2.0$        & $25.5\pm2.7$         & $25.7\pm1.5$         & $25.4\pm2.1$       \\ 
$\omega_{\mathrm Be}$ [$^\circ$]  & $90$                & $90$                 & $111.2\pm6.6$        & $111.9\pm9.0$      \\
$K_{\rm Be}$ [\kms]               & $3.9\pm0.7$\tablefootmark{a}         & $3.87\pm0.26$        & $3.9\pm0.7$\tablefootmark{a}          & $3.90\pm0.27$      \\
$K_{\rm pre-sd}$ [\kms]           & $60.55\pm0.68$      & $60.72\pm0.90$       & $61.07\pm0.66$       & $61.15\pm0.88$     \\
$\gamma$ [\kms]                   & $7.02\pm0.50$       & $7.56\pm0.19$        & $7.28\pm0.49$        & $7.57\pm0.19$        \\
\hline
$a$ [AU]                          & $0.3728\pm0.0086$   & $0.374\pm0.011$      & $0.3794\pm0.0079$    & $0.3800\pm0.0093$      \\
$q=M_{\rm pre-sd}/M_{\rm Be}$     & $0.064\pm0.012$     & $0.0638\pm0.0045$    & $0.064\pm0.012$      & $0.0638\pm0.0044$    \\
$M_{\rm total}$ [\Msun]           & $4.26\pm0.30$       & $4.28\pm0.36$        & $4.49\pm0.28$        & $4.51\pm0.33$      \\
$M_{\rm Be}$ [\Msun]              & $4.00\pm0.26$       & $4.03\pm0.34$        & $4.22\pm0.25$        & $4.24\pm0.31$      \\
$M_{\rm pre-sd}$ [\Msun]          & $0.258\pm0.054$     & $0.270\pm0.056$      & $0.270\pm0.056$      & $0.270\pm0.027$    \\
$d$ [pc]                          & $287.3\pm7.2$       & $287.6\pm8.9$        & $295.6\pm6.6$        & $296.0\pm8.0$      \\
\hline
rms astrometry [$\mu$as] & 25 & 25 & 19 & 19 \\
rms RV$_{\rm Be}$ [\kms]  & 1.64 & 1.53 & 1.56 & 1.53 \\
rms RV$_{\rm pre-sd}$ [\kms] & 4.92 & 4.96 & 4.83 & 4.85 \\
\hline
\end{tabular}
\end{center}
\tablefoot{The relative astrometry (astro) is given in Table~\ref{tab:astrometry}. The pre-sd RVs and $K_{\rm Be}$ were taken from \citet{2020ApJ...898L..44G}. $P$ is the orbital period, $T_0$ is the epoch of periastron or the epoch of the superior conjunction, $e$ is the eccentricity, $a''$ is the angular semimajor axis of the orbit, $i$ is the orbital inclination, $\Omega$ is the longitude of the ascending node, $\omega$ is the longitude of the periastron (fixed at $90^{\circ}$ for circular orbits), $K$ are the velocity semiamplitudes for the two components, $\gamma$ is the systemic velocity, $a$ is the orbital semimajor axis, $q=M_{\rm sdOB} / M_{\rm Be}$ is the mass ratio, $M_\mathrm{total}$ is the total mass, and $M_{\rm Be}$ and $M_{\rm sdOB}$ are the dynamical masses for the two components. ``rms'' stands for the rms residuals of specified datasets.\\
\tablefoottext{a}{Taken from \citet{2020ApJ...898L..44G}}
}
\end{table*}

%

\section{Discussion and conclusions}\label{sec:discussion}

Owing to the high spatial and spectral resolution and the overall good quality of the VLTI/GRAVITY data, we were able to obtain the astrometric orbit of the binary HR~6819, composed of a newly formed classical Be star and a bloated, recently stripped pre-sd companion. Previously published RV measurements enabled us to obtain a combined three-dimensional orbital solution, leading to the first precise model- and geometrical parallax-independent dynamical masses for this rare type of astrophysical object, which represents a very short stage in the evolution of interacting intermediate-mass binaries. Our orbital analysis indicates that the distance derived from the \textit{Gaia} DR3 parallax is biased due to the binary motions; the distance of $296.0\pm8.0$\,pc derived here is significantly lower than the \textit{Gaia} one \citep[$368\pm16$\,pc;][]{2021AJ....161..147B} and agrees well with the estimate of $310\pm60$\,pc based on SED fitting \citep{2020A&A...637L...3R}. The orbit appears to be slightly eccentric, with $e=0.0289\pm0.0058$, and thus for the final values of the distance and the dynamical masses listed in this section and in the abstract we take only the eccentric solutions into account (Table~\ref{tab:combined_solution}).

The dynamical mass of the Be star ($M_{\rm Be}=4.24\pm0.31$\,{\Msun}) is lower than what was estimated in recent works based on spectroscopy \citep{2020A&A...641A..43B,2021MNRAS.502.3436E}. The Be star may be hotter than expected; while \citet{2020A&A...641A..43B} assumed the Be star to have a spectral type of B2-3\,V from the fitted $T_{\rm eff}$, the dynamical mass is more consistent with B5\,V to B7\,V  \citep[following the mean parameters of main-sequence        dwarf stars based on the literature and catalog survey\footnote{\url{https://www.pas.rochester.edu/~emamajek/EEM_dwarf_UBVIJHK_colors_Teff.txt}};][version 2022.04.16]{2013ApJS..208....9P}. The Be star is rotating near-critically and is surrounded by a self-ejected Keplerian disk that is aligned with the binary orbit. These findings are consistent with the expectation that mass and angular momentum transfer from the originally more massive component is responsible for the rapid rotation of the Be star and the subsequent formation of its disk. The dynamical mass of the pre-sd star is at the lower end of the estimated error ranges from \citet{2020A&A...641A..43B} and \citet{2021MNRAS.502.3436E}. 

Of the known Be + (pre-)sdOB binaries, HR~6819 is only the second system with similarly well-constrained dynamical masses, after $\varphi$~Per \citep{2015A&A...577A..51M}, and only the ninth system with known dynamical masses when counting the seven parallax-dependent masses from \citet{2022ApJ...940...86K,2024ApJ...962...70K}. In addition, dynamical masses are available for OGLE-LMC-T2CEP-211, which is a post-mass-transfer eclipsing system composed of a likely classical Be star and a Type II Cepheid \citep{2018ApJ...868...30P}; this binary system might be in a very short progenitor stage of HR~6819-like systems when the stripped star is passing through an instability strip while evolving toward lower temperatures shortly after the mass transfer.

HR 6819's pre-sd star is expected to settle into a faint sdOB star within a few hundred thousand years \citep{2020A&A...641A..43B, 2021MNRAS.502.3436E}, at which point its spectrum will become dominated by the Be star. Compared to the known Be + sdOB systems with well-constrained parameters, HR~6819 has the most extreme mass ratio, the lowest stripped star mass, and the smallest orbital separation \citep{2015A&A...577A..51M,2024ApJ...962...70K}. It is most similar to the first confirmed Be + subdwarf B-type (sdB) system, $\kappa$~Dra, which has a slightly lower total mass but a $\sim$2 times higher mass ratio and a $\sim$30\% larger orbital separation \citep{2022ApJ...940...86K}. 

Given its similarity to $\kappa$~Dra, HR 6819's pre-sd star may become a relatively cool sdB star with a $T_{\rm eff}$ comparable to that of HR 6819's Be star, which would make it extremely hard to detect with current observing techniques, even compared to the already challenging sample of known Be + sdOB binaries. Using the estimated pre-sd radius of $\sim4.5$\,{\Rsun} \citep{2020A&A...641A..43B,2021MNRAS.502.3436E}, a future sdB radius of $\lesssim0.7$\,{\Rsun} \citep[sdB radius for $\kappa$~Dra;][]{2022ApJ...940...86K}, and a constant $T_{\rm eff}$ for simplicity, we can expect the pre-sd star's total flux contribution to decrease to $\lesssim1$\%. At the same angular separation, the higher contrast would put the stripped companion detection beyond the capabilities of VLTI/GRAVITY, and its spectral lines would not be visible in optical or far-UV spectra. Furthermore, the extreme mass ratio would make the measurement of the RV shifts of the Be star extremely elusive without prior knowledge of the orbit of the pre-sd companion, and the system would thus not be detectable as a spectroscopic binary. Only the low contrast resulting from the bloated evolutionary phase of the stripped star enabled both the angular resolution of its orbit with GRAVITY and the detection of the spectral lines of the future sdB star in the optical spectra. The angular resolution limit is less challenging for the CHARA array in the northern hemisphere, which has a longer maximum baseline than the VLTI (HR~6819 itself is not observable by CHARA).

It has been suggested by population synthesis studies that many, and possibly the majority of, Be stars are binary interaction products with faint evolved companions such as sdOBs or low-mass WDs \citep{1991A&A...241..419P,2014ApJ...796...37S,2021ApJ...908...67S}. Thus, if the configuration of HR~6819 (once the stripped star contracts) is common among Be stars, the important observational implication is that most present-day Be stars will appear as single. The relatively small number ($\approx$25) of confirmed descendants of HR~6819-like systems \citep[Be + sdOB binaries;][]{2018ApJ...853..156W, 2024ApJ...962...70K} might thus represent only the most easily detectable tip of the iceberg of this population. The incidence of spectroscopic Be binaries with low-mass companions of unknown nature \citep[e.g.,][]{2000ASPC..214..668G} as well as the prevalence of SED turndown among bright Be stars (see Sect.~\ref{sec:introduction}) are other observational clues suggesting that (mostly) invisible companions abound among Be stars.

As a by-product of the GRAVITY data analysis, we were able to measure the size of the Br$\gamma$ line-emitting region of the Be disk, which appears to be smaller than the orbit ($a''\sim1.3$\,mas) at FWHM$_{\mathrm{Br}\gamma\mathrm{, disk}} \sim 0.5$\,mas. On the other hand, the disk kinematics derived from the H$\alpha$ line suggest that the H$\alpha$ line-emitting region probably extends all the way to the pre-sd companion \citep{2020ApJ...898L..44G}. This is not completely unexpected, as the H$\alpha$ line emitting-region is generally the largest among all the emission lines. The angular size measurement of the Br$\gamma$ region should prove useful for future modeling of Be disk hydrodynamics in close binary systems \citep{rubio_et_al}, with a similar measurement already available for HR~2142 \citep{2024ApJ...962...70K}. 

While \citet{2020A&A...641A..43B} report a small emission peak in H$\alpha$ probably originating from circum-companion gas accreted from the outer parts of the Be star disk, we were not able to detect a similar signature in the Br$\gamma$ line. We did, however, detect weak signatures of the helium lines \spec{He}{i}{20587} and \spec{He}{i}{21126} that do originate from the pre-sd star, with \spec{He}{i}{20587} possibly showing a small level of emission (Sect.~\ref{sec:GRAVITY_RVs}, Fig.~\ref{fig:HeI20587}). This emission may originate from circum-companion gas accreted from the Be star disk. Thanks to the highly accurate wavelength calibration of our GRAVITY data ($<0.01$\%; see Fig.~\ref{fig:wavcal}), we could measure RVs from the two helium lines, which were found to agree well with the pre-sd RV curve based on FEROS spectra. This confirms that high-spectral-resolution GRAVITY data can be used to measure precise RVs, which may be extremely difficult to obtain by other means \citep[see also the case of HR~2142;][]{2024ApJ...962...70K}.

\begin{acknowledgements}

TS acknowledges support by the Israel Science Foundation (ISF) under grant number 0603225041.

\end{acknowledgements}

\bibliographystyle{aa} 
\bibliography{bibliography} 

\begin{appendix}

\onecolumn

\section{Detailed results from the fitting of GRAVITY data}

\begin{figure*}[]
\begin{center}
\includegraphics[width=17cm]{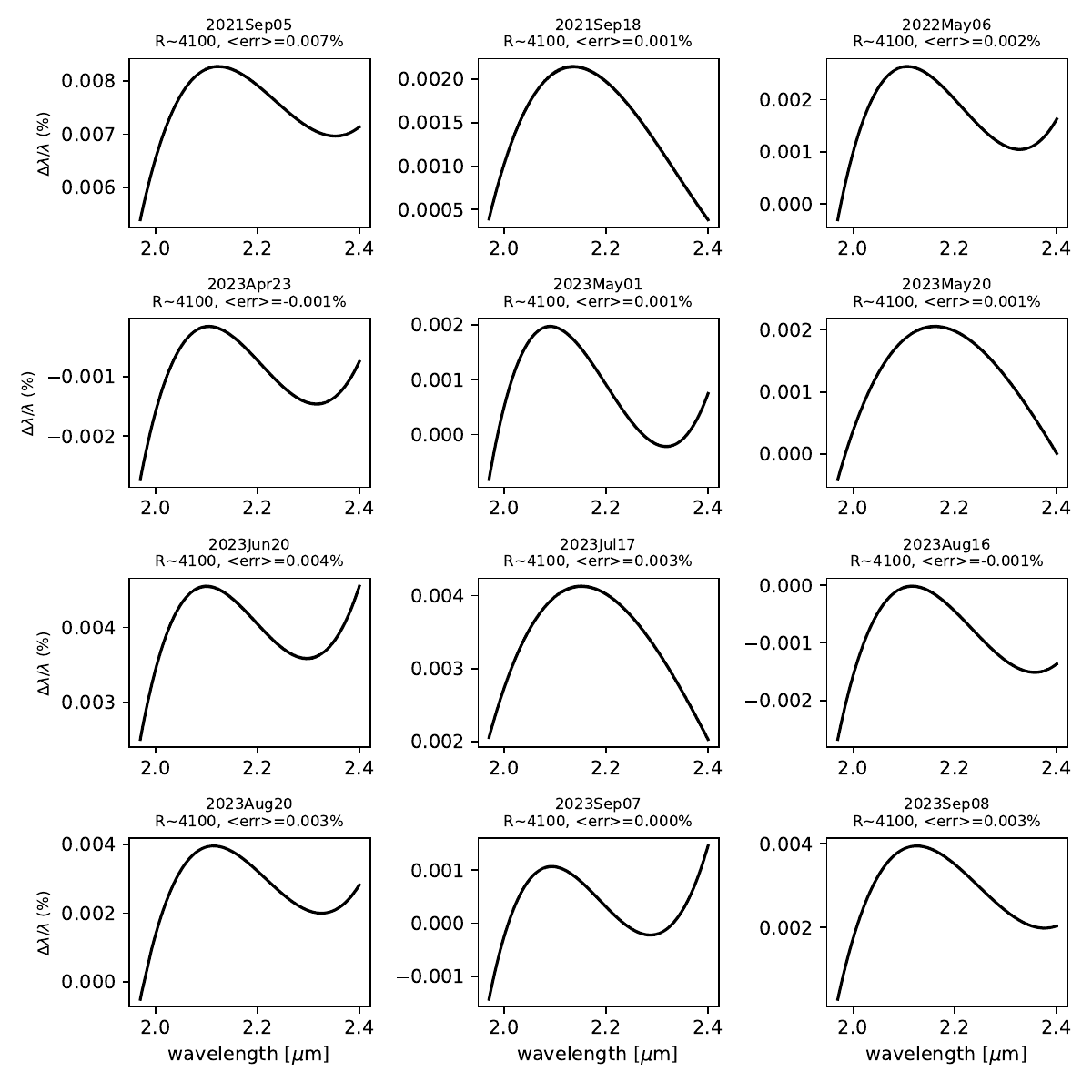}%
\end{center}
\caption[xx]{\label{fig:wavcal} Wavelength calibration accuracy, $\Delta \lambda / \lambda$, as a function of wavelength for each GRAVITY snapshot, as determined by PMOIRED fitting of telluric features. $R$ is the spectral resolution and <err> is the error in the wavelength calibration averaged across the full $K$-band range. An accuracy of 0.001\% in the $K$ band corresponds to $\sim3$\,\kms.
}
\end{figure*}

\begin{figure*}[]
\begin{center}
\includegraphics[width=17cm]{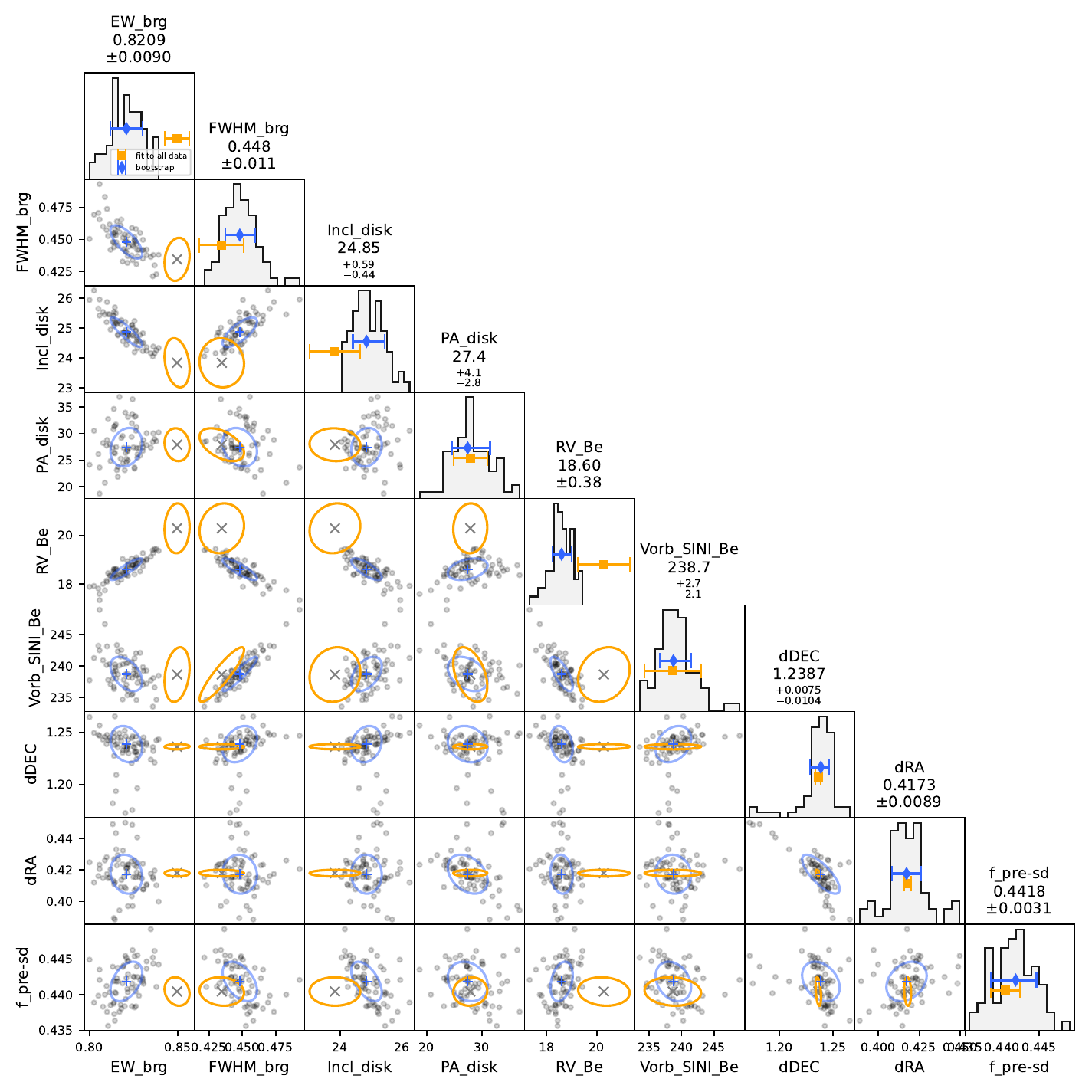}%
\end{center}
\caption[xx]{\label{fig:bootstrap} Probability distributions resulting from PMOIRED bootstrapping giving the final values of the model parameters for the 2023\,July\,17 snapshot. The quantities and units are the same as those shown in Tables~\ref{tab:astrometry} and \ref{tab:other_params} except for RV$_{\rm Be}$, which is displayed here before barycentric correction and before including the wavelength calibration error (which ends up dominating the uncertainty). Orange represents the original fit to the data, and blue shows the bootstrapping results.
}
\end{figure*}

\begin{figure*}[]
\begin{center}
\includegraphics[width=17cm]{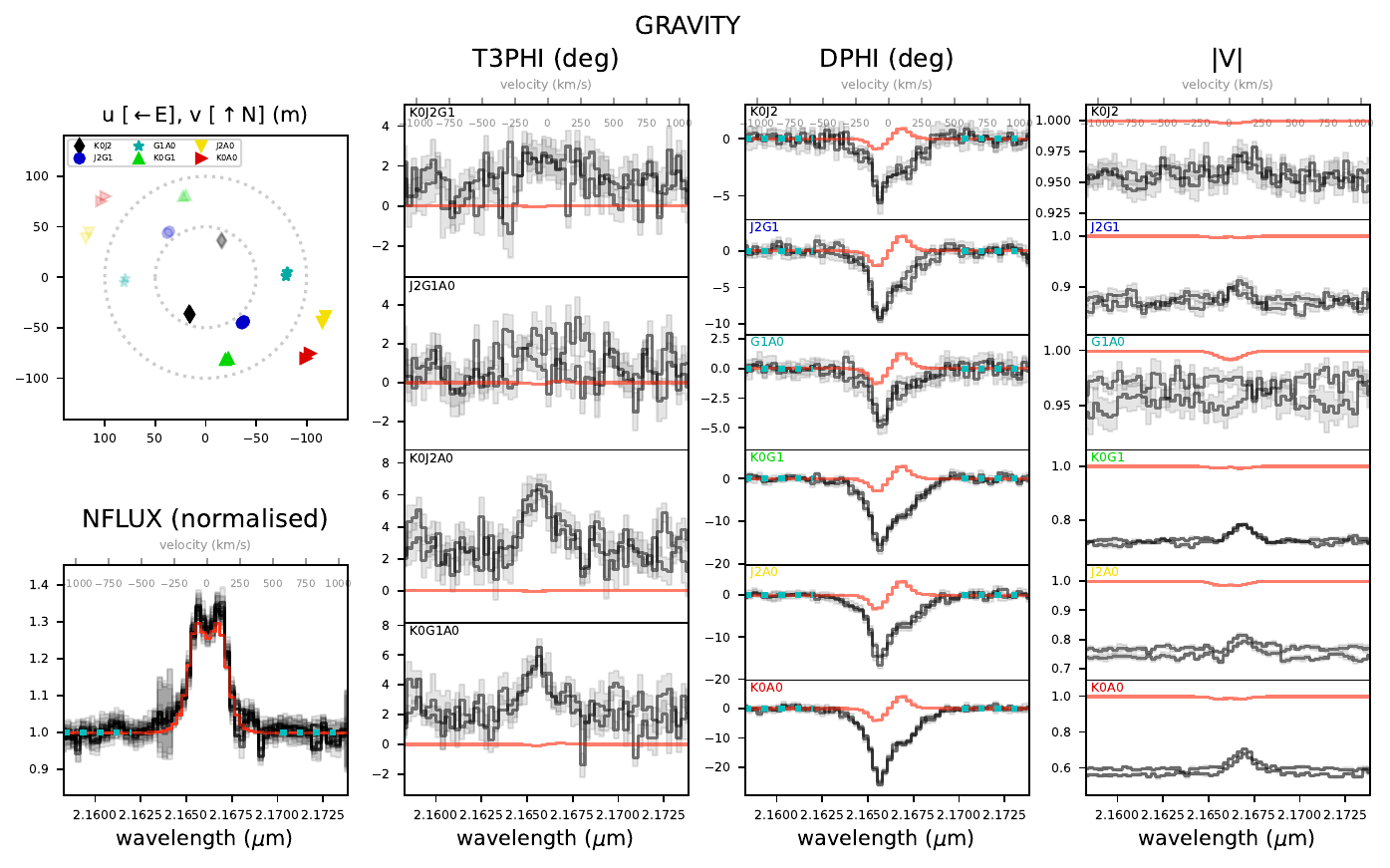}%
\end{center}
\caption[xx]{\label{fig:kepDisk_only} Same as Fig.~\ref{fig:bestfit}, but showing the geometrical model with the Keplerian disk parameters averaged across all epochs (Table~\ref{tab:other_params_average}) and without any contribution from the companion, i.e., as it would be expected from a single Be star.
}
\end{figure*}

\begin{figure*}[]
\begin{center}
\includegraphics[width=17cm]{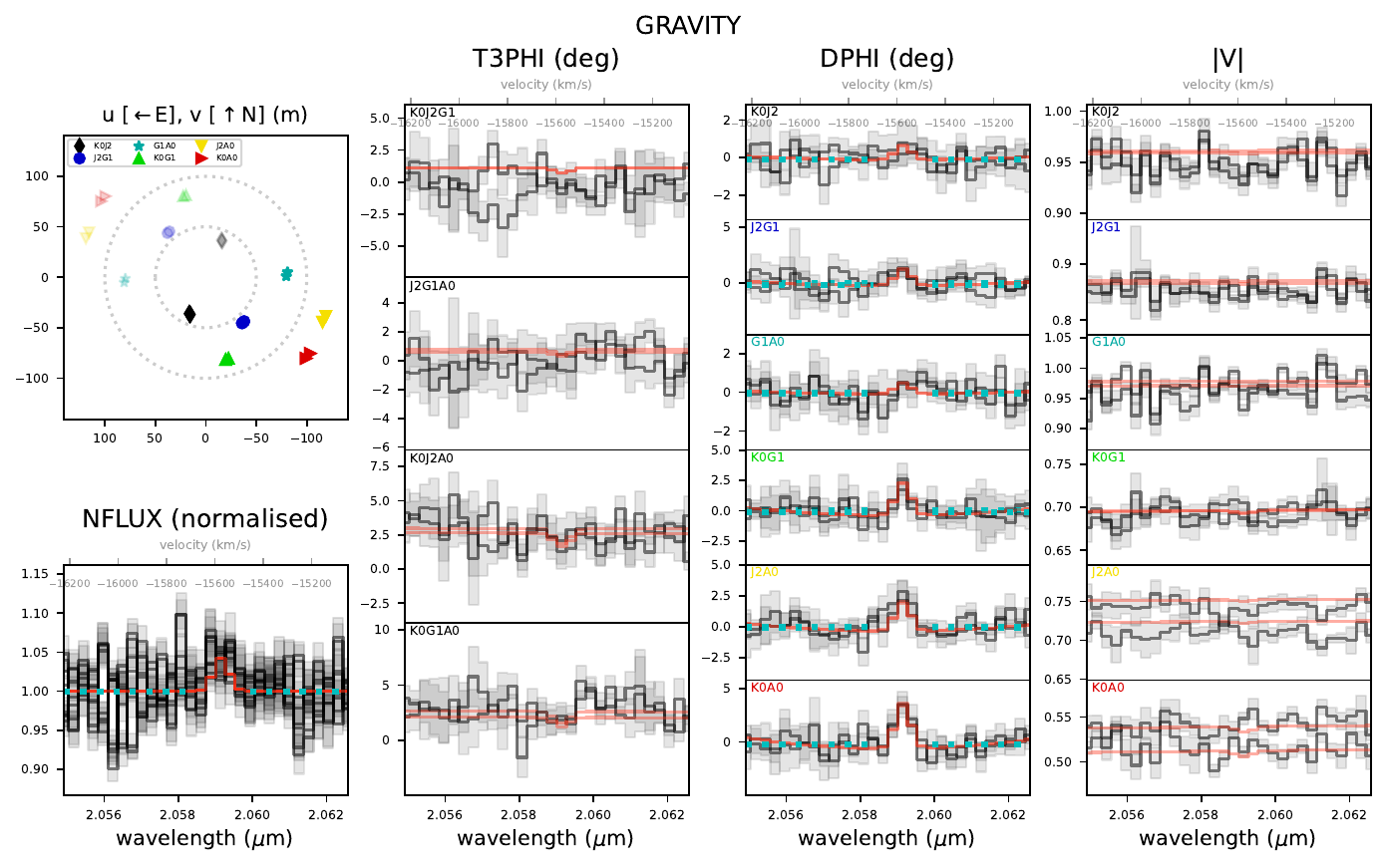}%
\end{center}
\caption[xx]{\label{fig:HeI20587} Same as Fig.~\ref{fig:bestfit}, but showing the spectral region around \spec{He}{i}{20587}. The line profile (NFLUX) shows a possible small emission component, while DPHI shows a clear deviation from zero in the spectral line. There is no discernible signal in T3PHI or |V|. Only DPHI was used for the fitting of the helium line RVs. 
}
\end{figure*}

The best-fit parameters are given for each epoch in Tables~\ref{tab:astrometry} and \ref{tab:other_params}. The wavelength calibration accuracy for each epoch is plotted in Fig.~\ref{fig:wavcal}. Bootstrapping results from the geometrical model fitting for a representative epoch are shown in Fig.~\ref{fig:bootstrap}. Averaged model for the Be star and its Keplerian disk as it would appear without a companion is compared to the dataset from the same epoch in Fig.~\ref{fig:kepDisk_only}. Finally, an example of the fit to the \spec{He}{i}{20587} line is shown in Fig.~\ref{fig:HeI20587}. 

\begin{table*}[h!]
\caption{Relative astrometric positions and flux ratios determined from the Br$\gamma$ region of GRAVITY data.}         
\label{tab:astrometry}      
\centering          
\begin{tabular}{c c c c c c c c c c c} 
\hline\hline       
Date & HJD$_\mathrm{mean}$ & Baselines & $\rho$ & PA & $\Delta$RA & $\Delta$Dec & $\sigma$-$a$ & $\sigma$-$b$ & $\sigma$-PA & $f_{\rm pre-sd, tot}$ \\ 
 & -2400000 & & [mas]  & [$^\circ$] & [mas] & [mas] & [mas] & [mas] & [$^\circ$] &  \\
\hline
2021\,Sep\,05 & 59463.625 & A0-G1-J2-J3 & 1.066 & 57.202 & 0.896 & 0.577 & 0.017 & 0.011 & 101.5 & 0.481$\pm$0.021\\
2021\,Sep\,18 & 59476.538 & D0-G2-J3-K0 & 1.072 & 192.472 & -0.231 & -1.047 & 0.048 & 0.028 & 162.0 & 0.540$\pm$0.028\\
2022\,May\,06 & 59706.711 & A0-G1-J2-J3 & 1.102 & 71.504 & 1.045 & 0.350 & 0.015 & 0.007 & 122.8 & 0.468$\pm$0.011\\
2023\,Apr\,23 & 60058.760 & A0-G1-J2-K0 & 1.134 & 350.192 & -0.193 & 1.117 & 0.049 & 0.014 & 127.0 & 0.429$\pm$0.015\\
2023\,May\,01 & 60066.747 & A0-G1-J2-K0 & 1.269 & 48.653 & 0.953 & 0.838 & 0.033 & 0.009 & 131.3 & 0.445$\pm$0.002\\
2023\,May\,20 & 60085.861 & A0-G1-J2-K0 & 1.245 & 223.548 & -0.858 & -0.902 & 0.024 & 0.011 & 5.0 & 0.429$\pm$0.006\\
2023\,Jun\,20 & 60116.635 & A0-G2-J2-J3 & 1.077 & 147.109 & 0.585 & -0.904 & 0.054 & 0.015 & 119.5 & 0.376$\pm$0.018\\
2023\,Jul\,17 & 60143.570 & A0-G1-J2-K0 & 1.307 & 18.601 & 0.417 & 1.239 & 0.014 & 0.007 & 137.5 & 0.442$\pm$0.003\\
2023\,Aug\,16 & 60173.624 & A0-G2-J2-J3 & 1.017 & 289.855 & -0.957 & 0.345 & 0.010 & 0.004 & 13.2 & 0.411$\pm$0.007\\
2023\,Aug\,20 & 60177.572 & A0-G1-J2-K0 & 1.138 & 330.386 & -0.562 & 0.989 & 0.044 & 0.011 & 8.7 & 0.447$\pm$0.014\\
2023\,Sep\,07 & 60195.505 & A0-G1-J2-K0 & 0.961 & 123.769 & 0.798 & -0.534 & 0.027 & 0.006 & 0.1 & 0.423$\pm$0.005\\
2023\,Sep\,08 & 60196.598 & A0-G1-J2-K0 & 0.992 & 137.460 & 0.670 & -0.731 & 0.051 & 0.009 & 30.5 & 0.400$\pm$0.013\\
\hline
\end{tabular}
\tablefoot{$\rho$ is the angular separation between the components, PA is the position angle (from north to east), $\Delta$RA and $\Delta$Dec are the companion coordinates relative to the primary in the east and north direction, respectively, $\sigma$-$a$ and $\sigma$-$b$ are the major and minor axes of the error ellipse, respectively, $\sigma$-PA is the position angle of the error ellipse (from north to east), and $f_{\rm pre-sd, tot}$ is the total flux fraction of the pre-sd star.\\
\tablefoottext{a}{The |V| observable had to be discarded from this dataset due to a miscalibration issue (|V|$>1.0$)}
}
\end{table*}

\begin{table*}[h!]
\caption[xx]{\label{tab:other_params} Parameters of the Be star and its Keplerian disk derived from GRAVITY data.}
\begin{center}
\begin{tabular}{lccccccc}
\hline\hline
Date & HJD$_\mathrm{mean}$ & $v_{\rm orb} \sin{i_{\rm disk}}$ & RV$_{\rm Be, bary}$ & $i_{\rm disk}$ & PA$_{\rm disk}$ & EW$_{\mathrm{Br}\gamma\mathrm{, disk}}$ & FWHM$_{\mathrm{Br}\gamma\mathrm{, disk}}$ \\ 
 & -2400000 & [mas]  & [\kms] & [$^\circ$] & [$^\circ$] & [nm] & [mas]  \\
 \hline
2021\,Sep\,05   & 59463.625     & $243.2\pm5.7$          & $-21\pm21$   & $25.67\pm0.31$  & $17.4\pm5.8$  & $0.7664\pm0.0025$     & $0.455\pm0.025$       \\
2021\,Sep\,18   & 59476.538     & $256.9\pm5.9$          & $15\pm12$    & $39.8\pm1.1$        & $27.4\pm2.2$      & $0.7142\pm0.0015$     & $0.516\pm0.025$       \\
2022\,May\,06   & 59706.711     & $257.4\pm5.7$          & $-3\pm12$    & $36.8\pm1.9$        & $30.5\pm2.4$      & $0.790\pm0.011$       & $0.488\pm0.020$       \\
2023\,Apr\,23   & 60058.760     & $230.0\pm5.2$          & $8\pm12$     & $38.1\pm2.0$        & $38.0\pm3.2$      & $0.6294\pm0.0071$     & $0.479\pm0.021$       \\
2023\,May\,01   & 60066.747     & $243\pm19$             & $4\pm12$     & $39.4$\tablefootmark{a}                 & $11.2\pm7.8$  & $0.5573\pm0.0012$     & $0.534\pm0.087$ \\
2023\,May\,20   & 60085.861     & $241.5\pm4.3$          & $23\pm12$    & $32.2\pm1.1$            & $24.1\pm1.1$  & $0.6292\pm0.0067$     & $0.530\pm0.017$       \\
2023\,Jun\,20   & 60116.635     & $294.8\pm7.9$          & $14\pm12$    & $36.39\pm0.36$  & $61.0\pm5.7$  & $0.7172\pm0.0005$     & $0.684\pm0.036$       \\
2023\,Jul\,17   & 60143.570     & $238.7\pm2.4$          & $9\pm12$     & $24.85\pm0.53$  & $27.4\pm3.5$  & $0.8209\pm0.0090$     & $0.448\pm0.011$       \\
2023\,Aug\,16   & 60173.624     & $236.5\pm4.7$          & $13\pm12$    & $34.1\pm1.4$        & $28.5\pm1.9$      & $0.7323\pm0.0069$     & $0.493\pm0.025$       \\
2023\,Aug\,20   & 60177.572     & $249.9\pm8.2$          & $6\pm12$     & $31.0\pm1.4$        & $35.9\pm2.4$      & $0.7608\pm0.0086$     & $0.523\pm0.037$       \\
2023\,Sep\,07   & 60195.505     & $234.6\pm4.6$          & $7\pm12$     & $21.9\pm1.1$        & $28.2\pm1.8$      & $0.979\pm0.020$       & $0.451\pm0.025$       \\
2023\,Sep\,08   & 60196.598     & $244\pm12$         & $17\pm12$        & $20.43\pm0.75$  & $27.3\pm1.7$  & $0.920\pm0.011$       & $0.420\pm0.045$       \\
\hline
\end{tabular}
\tablefoot{$v_{\rm orb} \sin{i_{\rm disk}}$ is the orbital velocity at the base of the Keplerian disk multiplied by the sine of the disk inclination, RV$_{\rm Be, bary}$ is the barycentric RV of the Be star, $i_{\rm disk}$ is the inclination of the circumstellar disk of the Be star, PA$_{\rm disk}$ is the position angle of the Be star disk measured from north to east, EW$_{\mathrm{Br}\gamma\mathrm{, disk}}$ is the equivalent width of the Br$\gamma$ emission line originating from the Be star disk, and FWHM$_{\mathrm{Br}\gamma\mathrm{, disk}}$ is the full width at half-maximum of the Gaussian representing the geometrical extent of Br$\gamma$ line-emitting region of the Be star disk.\\
\tablefoottext{a}{$i_{\rm disk}$ was not converging properly for this epoch (which lacks |V| data) and was therefore fixed at the orbital inclination taken from the eccentric solutions (Table~\ref{tab:combined_solution}).}
}
\end{center}
\end{table*}

\end{appendix}

\end{document}